\documentclass[aps,physrev,twocolumn,groupedaddress]{revtex4-2}

\usepackage{graphicx}%
\usepackage{gensymb}
\usepackage{float}

\begin{document}


\title{Photonic Weyl Points and Fermi Arc Surface States in a Self-Assembling Zinc-Blende Colloidal Crystal}


\author{Johnathon P. Gales}
\affiliation{Department of Physics, The Pennsylvania State University, University Park, PA, USA}

\author{Hengbin Cheng}
\affiliation{Department of Physics, The Pennsylvania State University, University Park, PA, USA}

\author{David J. Pine}
\affiliation{Department of Physics, New York University, New York, NY, USA}
\affiliation{Department of Chemical and Biomolecular Engineering, New York University, Brooklyn, NY, USA}

\author{Mikael C. Rechtsman}
\affiliation{Department of Physics, The Pennsylvania State University, University Park, PA, USA}


\date{\today}

\begin{abstract}
We characterize the design, photonic band structure, and self-assembly of a colloidal crystal capable of supporting robust topological Weyl points and Fermi arc surface states in 3D.
To achieve this, we propose a new zinc-blende structure, built off of recent progress in self-assembling diamond photonic crystals, which lacks inversion symmetry and supports photonic Weyl points. 
Furthermore, we show that the geometry can be optimized to make the Weyl point and its Fermi arc surface states experimentally observable in the photonic crystal's projected band structure. 
Finally, we perform molecular dynamics simulations to demonstrate that the geometry we propose for observing Weyl points is capable of being self-assembled with realistic interparticle interactions. 
Together, these results provide a platform for the assembly of large-scale photonic crystals supporting topological degeneracies and robust Fermi arc surface states in the visible and near-infrared.

\end{abstract}


\maketitle

\label{sec1}

The electromagnetic modes of photonic crystals may carry non-trivial topological properties that directly affect how light propagates within them~\cite{Ozawa2019TP}.  
In two dimensions, for example, photonic Chern insulators support robust edge states that do not scatter in the presence of disorder~\cite{raghu2008analogs, wang2009observation, rechtsman2013photonic,hafezi2013imaging, jin2025towards}.  
In three dimensions, perhaps the simplest and most robust form of topological photonic feature is a Weyl point, a two band linear degeneracy at a point in momentum space which is a source of Berry curvature~\cite{berry1984quantal, armitage2018weyl,lu2013weyl, noh2017experimental,vaidya2020observation,jorg2022observation}. 
The Weyl point degeneracies are only permitted in systems that lack parity-inversion-time-reversal (PT) symmetry.
When either inversion or time reversal symmetry is broken the structure is capable of supporting Weyl points, which come in pairs of opposite topological charges (integrated Berry flux), referred to as Chern numbers. 
Because the Chern number of a Weyl point is a quantized topological invariant, it cannot change under smooth perturbations to the lattice; the degeneracy can only be removed if oppositely charged degeneracies merge and annihilate~\cite{armitage2018weyl, Sarang2018Weyl}.
Analogous to the edge states associated with a 2D Dirac point, in graphene for example~\cite{Castro2009TheGraphene}, a crystal with Weyl points can support Fermi arc states on the surface of the crystal.
Due to these unique properties, Weyl points have been studied and observed in a variety of systems, including electrons in solids~\cite{Xu2015DiscoveryScience, Lv2015Observation} and photons in photonic crystals and waveguide arrays~\cite{lu2013weyl, noh2017experimental, vaidya2020observation, jorg2022observation}.

Studying Weyl points in 3D photonic crystals presents key challenges. 
In the optical regime, magneto-optical effects are too weak to break time-reversal symmetry appreciably, so instead lattices must be fabricated which lack inversion symmetry.
At technologically important visible and telecommunications wavelengths, current 3D photonic crystal fabrication methods are highly limited, both in spatial resolution and ability to scale to large system sizes.
Recently, Vaidya et al~\cite{vaidya2020observation,jorg2022observation} were able to fabricate a chiral woodpile structure, which lacks inversion symmetry, and observe its charge-2 quadratic Weyl point.
However, the wavelength (infrared) and system size (mm-scale) were both restricted by the top-down nanofabrication process.
The self-assembly and subsequent straining of diblock copolymer double gyroid structures has also been proposed as a method for achieving a Weyl point \cite{fruchart2018soft}, but the typical lattice constants for these assemblies \cite{hajduk1994gyroid} limit them to shorter ultraviolet wavelengths.
Here, we propose the use of colloidal self-assembly to realize a new 3D photonic crystal, based on the zinc-blende structure, with observable Fermi arc surface states at telecom and visible wavelengths.
The use of self-assembly, rather than top-down lithographic fabrication, immediately overcomes the fundamental challenges associated with wavelength and device scale in the realization of topological states of light.  

For nearly 40 years, colloidal self-assembly has been studied as a method for growing 3D photonic crystals \cite{pusey_phase_1986}, with a great deal of progress made in the pursuit of a complete photonic band gap \cite{yablonovitch_inhibited_1987, John1987Strong} in FCC \cite{Yablonovitch1993Photonic, John1987Strong, vlasov2001chip} and diamond~\cite{Ho1990,he2020colloidal} colloidal assemblies. 
Despite all of this progress, observing a Weyl point in a self-assembled colloidal crystal poses an interesting problem at optical scales. How do we break inversion symmetry?
Binary nanoparticle lattices with broken inversion symmetry have been assembled \cite{kalsin2006electrostatic,liu2016diamond}, but are inappropriate for achieving optical Weyl points due to their size and material absoption.
To address this, we build on the design of the diamond crystal assembled by He et al\cite{he2020colloidal}. 
By varying the particles in the diamond's 2-particle basis, a new zinc-blende crystal structure is formed. 
The zinc-blende crystal, a member of symmetry space group 216, lacks inversion symmetry and thus has the potential to support Weyl points and Fermi arc surface states.  
In detailed frequency and time domain numerical simulations, we establish the presence of such a Weyl point, as shown below.

\begin{figure*}[!htbp]
\centering
\includegraphics[width = \textwidth]{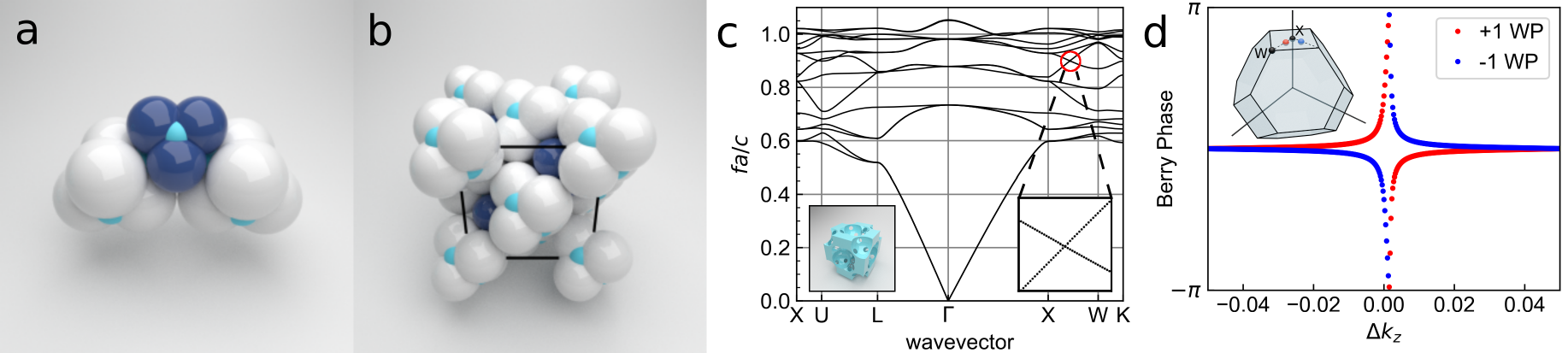}
\caption{A zinc-blende crystal comprised of large and small tetrahedral lobed patchy (TLPP) particles, with the ratio of the sizes $\phi = 1.35$. (a) A render of 2 large and 1 small particle, illustrating their bonding. (b) The cubic unit cell of the zinc-blende structure. (c)  Photonic band structure for the inverse zinc-blende of overlapping TLPP air holes in silicon ($n = 3.4$). A Weyl point degeneracy, circled in red, of the 8th and 9th bands lies on the X-W high symmetry line. Inset, a render of the unit cell for this inverse structure. (d) The Berry phase accumulated around different cross-sections of a sphere centered at the Weyl point. The Berry phase modulo $2\pi$ winds once around the Weyl point, confirming that the point has Chern number $C = +1$ (red). Its counterpart winds the oppposite direction confirming that it has Chern number $C = -1$ (blue). Inset, a depiction of the Brillouin zone, highlighting the positions of the $+1$ and $-1$ Weyl points, red and blue respectively.}
\label{fig:czb_strucbands}
\end{figure*}

A diamond crystal is an FCC Bravais lattice with a two-particle basis, where the two particles are identical.
If the two particles are not identical, then the diamond's inversion symmetry is broken: this is the zinc-blende crystal \cite{bragg1913structure}.
To self-assemble a zinc-blende structure from colloidal particles, we start from a diamond structure of tetrahedrally lobed patchy particles (TLPPs) \cite{wang2012colloids,he2020colloidal}.
The geometry of the colloids, and thus the final structure of the diamond \cite{gales2025crystallization}, is defined by the compression of the lobes into one another and the size of the patches on the faces of the tetrahedra.
While the patch size plays an important role in the assembly process \cite{gales2025crystallization}, the compression of the particles has the greatest impact on the photonic band structure of the assembled crystal.
For consistency with previous work, we quantify the patch size by the ratio of the patch extent $d_p$ to the radius of the spherical lobes $r_0$.
Similarly, the compression ratio of the lobes is defined to be $d_{cc}/2r_0$, where $d_{cc}$ is the center-to-center distance between neighboring lobes of the tetrahedron.
Supplementary Fig.~1 depicts the characterization of these two quantities.

To go from diamond to zinc-blende, we simply take the two identical TLPPs in the diamond basis with compression $d_{cc}/2r_0 = 0.70$ and make one bigger, with lobe radius $r_l$, and the other smaller, with lobe radius $r_s$.
At a given compression, if the size difference between the two TLPPs becomes too great, then the smaller particles are geometrically incapable of forming all four of their bonds due to steric exclusion.
Taking this into account, we choose the ratio of the particle sizes to be $\phi = r_l/r_s = 1.35$.
As seen in Fig.~\ref{fig:czb_strucbands}a, the small particle (blue) is able to bond with multiple large particles (white) without their lobes overlapping.
Figure~\ref{fig:czb_strucbands}b depicts the cubic unit cell of this zinc-blende structure, illustrating that it is still an FCC lattice with a basis of one large particle (white) and one small particle (blue).


In current synthesis methods, the TLPPs comprise four polystyrene spheres with four TPM patches. 
Both of these materials are dielectrics with a low refractive index, approximately 1.6 for polystyrene and 1.5 for TPM. 
These refractive indices do not provide sufficient contrast with air to open gaps between the bands, where the Weyl point can be observed.

For experimental characterization of the Weyl point, it is important that the degeneracy be well separated from the other bands for as large a range of wavevectors as possible. 
To spread the bands out and reveal the Weyl point band crossing clearly, we first increase the index of the lattice by infiltrating the colloidal assembly with silicon. 
This silicon infiltration process was previously carried out with low-pressure chemical vapor deposition to achieve a band gap in the FCC lattice \cite{vlasov2001chip}.
After infiltration, the zinc-blende crystal is inverted: it comprises TLPP-shaped air holes embedded in a matrix of high-index silicon, with refractive index $n_{Si} = 3.4$.
Additionally, we optimize the band structure by increasing the size of the air holes to make neighboring holes overlap. 
This change further optimizes the band structure, as was done for the diamond of spheres \cite{Ho1990} and TLPPs \cite{he2020colloidal}. 
In experiment, this can be achieved by growing additional sacrificial material on the particles before silicon inversion. 
We find that a sacrificial layer of thickness $t = 0.05r_l$ is optimal for the present zinc-blende structure. 

The photonic band structure of this inverse zinc-blende is calculated using the plane wave expansion method with the publicly available software MIT photonic bands (MPB)\cite{johnson2001block} and is plotted in Fig.~\ref{fig:czb_strucbands}c. 
A Weyl point is observed in the band structure in the form of a linear crossing, circle in red and shown in the inset, between the 8th and 9th bands along the $X$-$W$ high symmetry line at the Brillouin zone boundary.
Indeed, the $X$-$W$ line of zinc-blende has been predicted to support charge-1 Weyl points by a generic symmetry-enforced $k\cdot p$ analysis, where the Weyl points are pinned to the $X-W$ line by a 2-fold rotation symmetry~\cite{yu2022encyclopedia}.
The lattice of non-overlapping air holes, shown in Supplementary Fig.~2, also has a Weyl point, but the gap it sits in is not nearly as omnidirectional as for the overlapping air holes.

To confirm that the degeneracy (circled in red) between the 8th and 9th bands has a non-zero topological charge and is therefore a Weyl point, we use MPB to determine the Chern number of the degeneracy. 
To do so, we use the magnetic field eigenmodes from MPB and the algorithm first proposed by Resta\cite{resta2000manifestations} to calculate the associated Berry phase across the degeneracy. 
In short, the magnetic field eigenmodes of the 8th band are calculated along a closed loop in the Brillouin zone.
The fields are then used to calculate the Berry phase accrued around that loop.
When the loops are chosen to be circular cross-sections of a sphere centered on the degeneracy, each with a different $k_z$, the Berry phase acquired is calculated as a function of the $k_z$ of the loops \cite{vaidya2020observation}. 
For a topologically non-trivial degeneracy, the Berry phase as a function of $k_z$ will wind (modulo $2\pi$) as the circular loops sweep from one pole of the sphere to the other; the number of times the Berry phase winds is the topological charge of the degeneracy.
Figure~\ref{fig:czb_strucbands}d shows the Berry phase calculation for the inverse zinc-blende of overlapped spheres. 
For this structure, the Berry phase winds once, confirming that this degeneracy is a charge $C = +1$ Weyl point (red).

A system with broken inversion symmetry supports multiple Weyl points, which come in pairs.
The $C = +1$ Weyl point must also come with an oppositely charged $C = -1$ degeneracy, whose position in the Brillouin zone is dictated by the symmetries of the structure.
In this case, the zinc-blende structure's mirror plane along the line $x = y$ ensures that the $C = +1$ Weyl point on the $k_x$ axis is paired up with a $C = -1$ Weyl point at the same position on the $k_y$ axis, see \ref{fig:czb_strucbands}d Inset.
This Weyl point's existence and charge are confirmed by performing the same Berry phase calculation at this position on the $k_y$ line.
At this position, the Berry phase winds once in the opposite direction, confirming that this is a charge $C = -1$ Weyl point.
There is another pair of Weyl points on the same $k_z$ plane along the $-k_x$ and $-k_y$ lines, with $C = +1$ and $C = -1$ respectively.
%


While the zinc-blende structure's Weyl point is clearly resolvable in the 3D band structure in Fig.~\ref{fig:czb_strucbands}, current experimental techniques are incapable of directly measuring a full 3D photonic band structure in the optical regime.
For example, if the light is incident on the (001) face of the crystal, then Bloch's theorem is satisfied in the $x$ and $y$ directions but not the $z$ direction.
This means that incoming light that has some momentum $k = (k_x, k_y, k_z)$ must excite a mode of the crystal with the same $k_x$ and $k_y$, but can excite modes of any $k_z$.
Since, in general, all $k_z$ modes are excited for a particular $k_x$ and $k_y$, the eigenmodes of the crystal within the 3D Brillouin zone are all projected onto the 2D $k_z = 0$ plane of the Brillouin zone.
The projection of the band structure is unavoidable in common experimental characterization techniques such as angle-resolved FTIR (Fourier transform infrared spectroscopy) measurements~\cite{vaidya2020observation}.
Often, features of the 3D band structure become unresolvable in the projected band structure.
While the bands may be well separated at a particular wavevector, eigenmodes with the same frequency at a different $k_z$ can project on top of the band feature of interest.

\begin{figure}[!htbp]
    \centering
    \includegraphics[scale = 0.8]{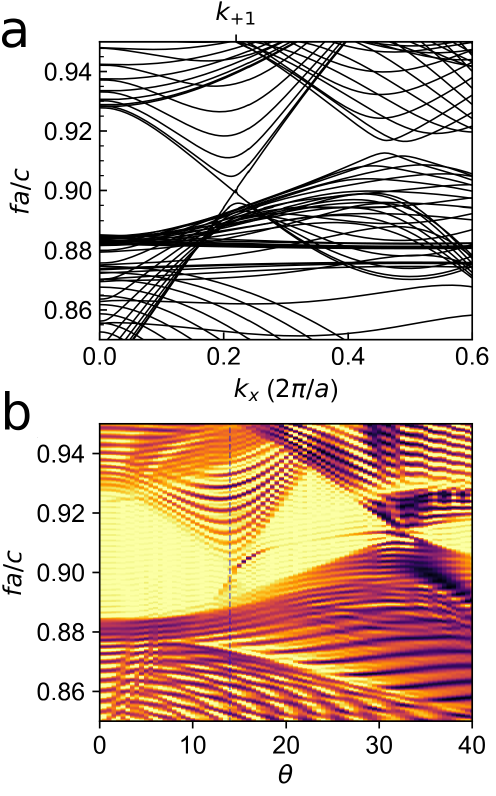}
    \caption{(a) Band structure (MPB) along $k_x$ projected down onto $k_z = 0$. The Weyl point crossing is observed in the stop-gap of the projected bands at $fa/c = 0.90$ and $k_x = k_{+1} = 0.22$. (b) FDTD angle resolved simulation (Tidy3D) of the projected band structure for a $20a$ thick zinc-blende along the $k_x$ line. The Weyl point is observed at frequency $fa/c = 0.90$ and angle $\theta = 14\degree$. The FDTD simulation exhibits modes in the gap (arrow), not present in the bulk projected bands. These are the Fermi arc surface states of the crystal.}
    \label{fig:czb_proj}
\end{figure}

We confirm that our zinc-blende structure's Weyl point is experimentally observable by calculating the projected band structure with MPB and simulating a typical angle-resolved transmission/reflection experiment using finite-difference time-domain (FDTD) simulations performed with the commercially available software package Tidy3D~\cite{tidy3dfdtd}.
In MPB, we calculate the modes along the $k_x$ axis as a function of $k_z$ from $k_z = 0$ to $k_z = 1$ ($2\pi/a$).
The projected band structure is then produced by plotting all of the different $k_z$ bands on top of one another, as shown in Fig.~\ref{fig:czb_proj}a.

\begin{figure*}[!htbp]
    \centering
    \includegraphics[width = \textwidth]{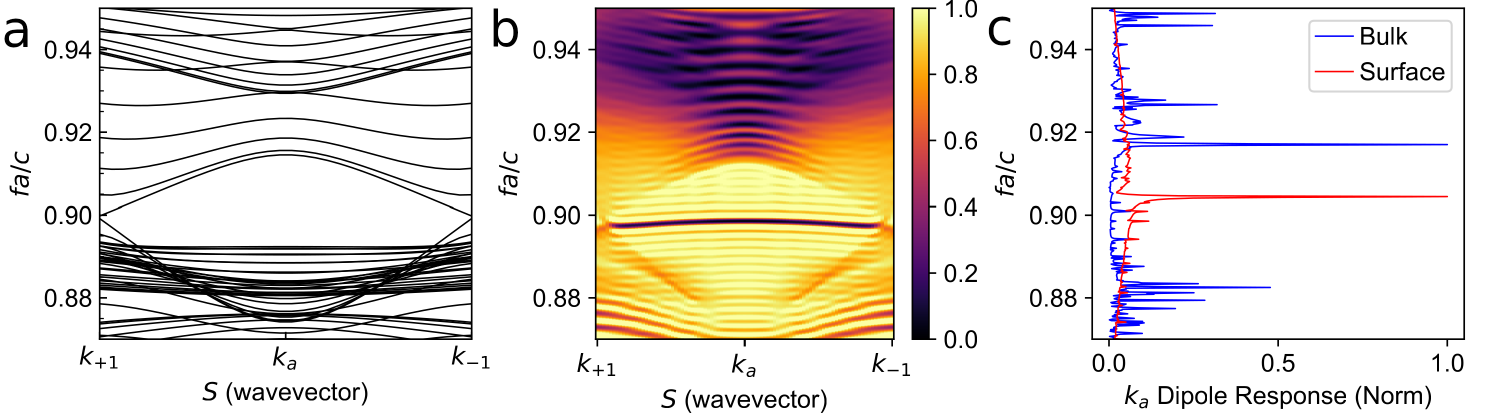}
    \caption{(a) The MPB projected bulk band structure along the straight line path $\mathit{S}$ from $k_{+}$ to $k_{-1}$. The bulk bands have a gap along this path between the two Weyl points.(b) The FDTD projected band structure along the straight line path $\mathit{S}$ from $k_{+1}$ to $k_{-1}$. There is a dark band present in the gap of the bulk bands, which connects the two Weyl points. This dark band is the Fermi arc surface state.(c) The FDTD frequency response at $k_a$ to dipoles placed in the bulk of the crystal (blue) or on the surface of the crystal (red). The bulk dipoles reproduce the bulk band gap from the previous plots, slightly shifted to higher frequency. The surface measurement has only one strong mode in the gap of the bulk bands, corresponding to the Fermi arc surface state.
    }
    \label{fig:czb_surf}
\end{figure*}

The Weyl point identified in Fig.~\ref{fig:czb_strucbands} lies on the $X$-$W$ high-symmetry line, on the edge of the Brillouin zone.
Figure~\ref{fig:czb_proj}a shows that the Weyl point projects down from $k_z = 1$ into a gap of the projected band structure where it can still be observed.
The separation of the Weyl point from the projected bands above and below it is a consequence of optimizing the size of the air holes with respect to the 3D band structure.

Figure~\ref{fig:czb_proj}b shows the projected band structure obtained from angle-resolved FDTD reflection simulations, detailed in Supplementeray Section 3.
These simulations emulate the expected results of an angle-resolved spectroscopy measurement, such as those previously performed with FTIR~\cite{vaidya2020observation,jorg2022observation}.
In these simulations, different wavevectors along the $k_x$ line are probed by varying the angle $\theta$ of the plane-wave source incident on the crystal's surface.
Since the crystal is finite in $Z$, we choose the surface termination to be a distance $0.025a$ inside the unit cell, seen in Supplementary Fig.~3.
This projected band structure agrees quite well with the frequency-domain calculation of MPB, confirming that the Weyl point, located at $\theta = 14\degree$, can be observed in a realistic experimental measurement.

Interestingly, there is one band feature in the FDTD spectrum, highlighted by the arrow in Fig.~\ref{fig:czb_proj}b, that is not found in the bulk photonic bands calculated by MPB.
Near the Weyl point and in the smaller bulk gap around $\theta = 32\degree$, there is an additional band that is not present in the bulk band structure in Fig.~\ref{fig:czb_proj}a.
Analyzing the fields in FDTD confirms that this mode is a surface state confined to the air-crystal interface, as seen in Supplementary Fig.~3.
Since the surface state emanates from the Weyl point, it is a topological Fermi arc surface state, which is analogous to the topological edge modes associated with a Dirac point in 2D~\cite{Castro2009TheGraphene}.
The wavevectors at which the surface modes in Fig.~\ref{fig:czb_proj}b disappear make it clear that this path along $k_x$ is not the most useful one for fully identifying the Fermi arc surface state.
A Fermi arc surface state runs from a $C = +1$ Weyl point to a $C = -1$ Weyl point\cite{armitage2018weyl}. 
To better resolve this surface state and confirm its topological nature, we measure the modes along a new path through the Brillouin zone, which connects $k_{+1}$ to $k_{-1}$.

We choose the path, denoted $\mathit{S}$, to be the straight line from $k_{+1}$, on the $k_x$ axis, to $k_{-1}$, on the $k_y$ axis, as depicted in Supplementary Fig.~4.
The bulk projected band structure for this path is calculated in MPB and plotted in Fig.~\ref{fig:czb_surf}a.

After confirming there is a bulk gap along the path $\mathit{S}$, we perform FDTD simulations along the same path to reveal the Fermi arc surface state in \ref{fig:czb_surf}b.
The Fermi arc surface state appears as a dark band in the gap connecting the $C = +1$ and $C = -1$ Weyl points.
The Fermi arc surface state's dispersion is nearly flat, which means that the path $\mathit{S}$ is nearly the iso-frequency contour of the Fermi arc surface state at $fa/c = 0.898$.
The topological nature of the state is further reinforced by confirming that this Fermi arc iso-frequency contour is unique to this Weyl point pair, as shown in Supplementary Fig.~5.

With the topological nature of the state confirmed, we next characterize the degree to which the state is confined to the crystal surface. 
As an initial check, we extract the mode profile from an FDTD simulation at wavevector $k_a$ and frequency $fa/c = 0.898$, plotted in Supplementary Fig.~6, to see that these fields are entirely confined to the incident surface. 
Furthermore, we run two new FDTD simulations, described in Supplementary Section 8, using point dipole sources to excite the modes at two different positions: the center of the bulk crystal, and the surface of the crystal.
The modes excited by the bulk dipoles, indicated in blue in \ref{fig:czb_surf}c, show a strong response on the top and bottom edges of the bulk band gap, but no response inside the gap where the Fermi arc surface state lives. 
The response to the surface dipoles, plotted in red in \ref{fig:czb_surf}c, is dominated by a high-Q response in the middle of the gap measured by the bulk dipoles. 
The strong peak confirms the localization of the Fermi arc state to the surface. 

The topological state at the air-crystal interface has a high Q-factor (and thus a narrow linewidth), as seen in both the darkness of the band in \ref{fig:czb_surf}b and the large peak in \ref{fig:czb_surf}c, making it much more feasible to observe.
Naively one may assume that a mode on the air-crystal interface would be highly leaky into the air, meaning its linewidth would be broad, making it difficult to observe.
In that case, a cladding would be required to confine the state, which would add additional difficulty to the assembly, infiltration, and characterization of such a structure.
The naturally strong resonance of our zinc-blende's Fermi arc surface state means that such a cladding is not necessary and the Fermi arc state can be observed straightforwardly.

Now that we have confirmed that the zinc-blende geometry discussed above has an observable Weyl point and Fermi arc surface states, we ensure that this geometry can be realistically self-assembled by performing molecular dynamics simulations. The molecular dynamics simulations, performed with HOOMD-Blue \cite{hoomd-blue}, allow for thousands of TLPPs to diffuse and sediment under gravity onto a surface while their patches and lobes interact via pair potentials. As in previous work studying TLPP diamond assembly\cite{gales2025crystallization}, we model DNA-coated patches\cite{he2020colloidal,wang_synthetic_2015} using a short-range Wang-Frenkel potential\cite{wang2020lennard}. This potential, described in Supplementary Section 9, was found to model pairwise DNA interactions very well in experimental characterizations\cite{cui2022comprehensive}. Further details of the simulations are described in Supplementary Section 10.

The particles in the molecular dynamics simulations have the same size ratio, $\phi = 1.35$, and compression ratio, $d_{cc}/2r_0 = 0.70$, that was used to produce the Weyl point in the previous sections.
While it does not have a strong effect on the band structure, the extent of the patches $d_p$ plays a crucial role in self-assembly. 
If the patches are too recessed, then the particles are inhibited from binding. 
If the patches are too extended, then the TLPP bonds are not restricted to the conformation necessary to produce diamond/zinc-blende\cite{gales2025crystallization}. 
For simplicity, we choose for the large and small TLPPs to have the same geometry, patch size ratio $d_p/r_l$ and compression ratio $d_{cc}/2r_0$, just scaled larger or smaller. 
After performing a number of simulations at different patch size ratios, we find that there is a range in which the particles crystallize and form the zinc-blende lattice.

\begin{figure}[H]
    \centering
    \includegraphics[scale = 0.65]{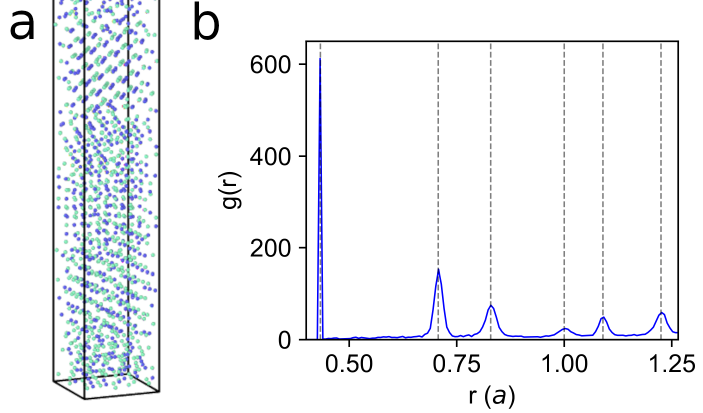}
    \caption{
    (a) Particle centers of mass from the final frame of the MD simulation. The particles have formed multiple zinc-blende crystals. 
    (b) Pair correlation function $g(r)$ calculated for the centers of mass of the final structure. The distinct $g(r)$ peaks coincide precisely with the positions of the 1st-6th nearest neighbors of a perfect zinc-blende (dashed lines).}
    \label{fig:czb_assembly}
\end{figure}

Figure~\ref{fig:czb_assembly} shows the results of the HOOMD molecular dynamics simulation for particles with patch size ratio $d_p/r_l = 1.38$. 
The order and crystalline stacking of these different grains becomes clearer when looking at only TLPP centers of mass in Fig.~\ref{fig:czb_assembly}a.
These centers of mass are used to calculate the particles' pair correlation function $g(r)$.
Plotted in Fig.~\ref{fig:czb_assembly}b, the distinct peaks in $g(r)$ confirm the crystallinity of the final structure assembled.
These peaks coincide with the neighbor distances of the zinc-blende lattice, confirming, along with the strict large-small binding, that the particles have self-assembled into the desired zinc-blende structure.

\vspace*{0.25cm}

We have proposed a realistically self-assemblable colloidal lattice that realizes charge-1 Weyl points by the breaking of inversion symmetry.
The results of photonic and molecular dynamics simulations confirm that self-assembly of a zinc-blende colloidal crystal is an experimentally feasible method for the realization of topologically protected Weyl points and Fermi arc surface states in the visible and near-infrared.
Furthermore, the Weyl points are found to be unobstructed by other bands and the Fermi arc surface states have high quality factor without a need for cladding, making them highly accessible experimentally.
This provides a clear path toward the experimental realization of colloidal Weyl points and topologically robust Fermi arc surface states.
More generally, we believe this work showcases the potential for studying topological photonic features in self-assembled colloidal photonic crystals.
These highly scalable systems could provide an excellent platform for the study of topologically non-trivial 3D photonics.

This work was supported by the Army Research Office under MURI Grant No. W911NF-24-1-0224.


\begin{thebibliography}{38}%
\makeatletter
\providecommand \@ifxundefined [1]{%
 \@ifx{#1\undefined}
}%
\providecommand \@ifnum [1]{%
 \ifnum #1\expandafter \@firstoftwo
 \else \expandafter \@secondoftwo
 \fi
}%
\providecommand \@ifx [1]{%
 \ifx #1\expandafter \@firstoftwo
 \else \expandafter \@secondoftwo
 \fi
}%
\providecommand \natexlab [1]{#1}%
\providecommand \enquote  [1]{``#1''}%
\providecommand \bibnamefont  [1]{#1}%
\providecommand \bibfnamefont [1]{#1}%
\providecommand \citenamefont [1]{#1}%
\providecommand \href@noop [0]{\@secondoftwo}%
\providecommand \href [0]{\begingroup \@sanitize@url \@href}%
\providecommand \@href[1]{\@@startlink{#1}\@@href}%
\providecommand \@@href[1]{\endgroup#1\@@endlink}%
\providecommand \@sanitize@url [0]{\catcode `\\12\catcode `\$12\catcode `\&12\catcode `\#12\catcode `\^12\catcode `\_12\catcode `\%12\relax}%
\providecommand \@@startlink[1]{}%
\providecommand \@@endlink[0]{}%
\providecommand \url  [0]{\begingroup\@sanitize@url \@url }%
\providecommand \@url [1]{\endgroup\@href {#1}{\urlprefix }}%
\providecommand \urlprefix  [0]{URL }%
\providecommand \Eprint [0]{\href }%
\providecommand \doibase [0]{https://doi.org/}%
\providecommand \selectlanguage [0]{\@gobble}%
\providecommand \bibinfo  [0]{\@secondoftwo}%
\providecommand \bibfield  [0]{\@secondoftwo}%
\providecommand \translation [1]{[#1]}%
\providecommand \BibitemOpen [0]{}%
\providecommand \bibitemStop [0]{}%
\providecommand \bibitemNoStop [0]{.\EOS\space}%
\providecommand \EOS [0]{\spacefactor3000\relax}%
\providecommand \BibitemShut  [1]{\csname bibitem#1\endcsname}%
\let\auto@bib@innerbib\@empty
\bibitem [{\citenamefont {Ozawa}\ \emph {et~al.}(2019)\citenamefont {Ozawa}, \citenamefont {Price}, \citenamefont {Amo}, \citenamefont {Goldman}, \citenamefont {Hafezi}, \citenamefont {Lu}, \citenamefont {Rechtsman}, \citenamefont {Schuster}, \citenamefont {Simon}, \citenamefont {Zilberberg},\ and\ \citenamefont {Carusotto}}]{Ozawa2019TP}%
  \BibitemOpen
  \bibfield  {author} {\bibinfo {author} {\bibfnamefont {T.}~\bibnamefont {Ozawa}}, \bibinfo {author} {\bibfnamefont {H.~M.}\ \bibnamefont {Price}}, \bibinfo {author} {\bibfnamefont {A.}~\bibnamefont {Amo}}, \bibinfo {author} {\bibfnamefont {N.}~\bibnamefont {Goldman}}, \bibinfo {author} {\bibfnamefont {M.}~\bibnamefont {Hafezi}}, \bibinfo {author} {\bibfnamefont {L.}~\bibnamefont {Lu}}, \bibinfo {author} {\bibfnamefont {M.~C.}\ \bibnamefont {Rechtsman}}, \bibinfo {author} {\bibfnamefont {D.}~\bibnamefont {Schuster}}, \bibinfo {author} {\bibfnamefont {J.}~\bibnamefont {Simon}}, \bibinfo {author} {\bibfnamefont {O.}~\bibnamefont {Zilberberg}},\ and\ \bibinfo {author} {\bibfnamefont {I.}~\bibnamefont {Carusotto}},\ }\bibfield  {title} {\bibinfo {title} {Topological photonics},\ }\href {https://doi.org/10.1103/RevModPhys.91.015006} {\bibfield  {journal} {\bibinfo  {journal} {Rev. Mod. Phys.}\ }\textbf {\bibinfo {volume} {91}},\ \bibinfo {pages} {015006} (\bibinfo {year} {2019})}\BibitemShut {NoStop}%
\bibitem [{\citenamefont {Raghu}\ and\ \citenamefont {Haldane}(2008)}]{raghu2008analogs}%
  \BibitemOpen
  \bibfield  {author} {\bibinfo {author} {\bibfnamefont {S.}~\bibnamefont {Raghu}}\ and\ \bibinfo {author} {\bibfnamefont {F.~D.~M.}\ \bibnamefont {Haldane}},\ }\bibfield  {title} {\bibinfo {title} {Analogs of quantum-hall-effect edge states in photonic crystals},\ }\href@noop {} {\bibfield  {journal} {\bibinfo  {journal} {Physical Review A—Atomic, Molecular, and Optical Physics}\ }\textbf {\bibinfo {volume} {78}},\ \bibinfo {pages} {033834} (\bibinfo {year} {2008})}\BibitemShut {NoStop}%
\bibitem [{\citenamefont {Wang}\ \emph {et~al.}(2009)\citenamefont {Wang}, \citenamefont {Chong}, \citenamefont {Joannopoulos},\ and\ \citenamefont {Solja{\v{c}}i{\'c}}}]{wang2009observation}%
  \BibitemOpen
  \bibfield  {author} {\bibinfo {author} {\bibfnamefont {Z.}~\bibnamefont {Wang}}, \bibinfo {author} {\bibfnamefont {Y.}~\bibnamefont {Chong}}, \bibinfo {author} {\bibfnamefont {J.~D.}\ \bibnamefont {Joannopoulos}},\ and\ \bibinfo {author} {\bibfnamefont {M.}~\bibnamefont {Solja{\v{c}}i{\'c}}},\ }\bibfield  {title} {\bibinfo {title} {Observation of unidirectional backscattering-immune topological electromagnetic states},\ }\href@noop {} {\bibfield  {journal} {\bibinfo  {journal} {Nature}\ }\textbf {\bibinfo {volume} {461}},\ \bibinfo {pages} {772} (\bibinfo {year} {2009})}\BibitemShut {NoStop}%
\bibitem [{\citenamefont {Rechtsman}\ \emph {et~al.}(2013)\citenamefont {Rechtsman}, \citenamefont {Zeuner}, \citenamefont {Plotnik}, \citenamefont {Lumer}, \citenamefont {Podolsky}, \citenamefont {Dreisow}, \citenamefont {Nolte}, \citenamefont {Segev},\ and\ \citenamefont {Szameit}}]{rechtsman2013photonic}%
  \BibitemOpen
  \bibfield  {author} {\bibinfo {author} {\bibfnamefont {M.~C.}\ \bibnamefont {Rechtsman}}, \bibinfo {author} {\bibfnamefont {J.~M.}\ \bibnamefont {Zeuner}}, \bibinfo {author} {\bibfnamefont {Y.}~\bibnamefont {Plotnik}}, \bibinfo {author} {\bibfnamefont {Y.}~\bibnamefont {Lumer}}, \bibinfo {author} {\bibfnamefont {D.}~\bibnamefont {Podolsky}}, \bibinfo {author} {\bibfnamefont {F.}~\bibnamefont {Dreisow}}, \bibinfo {author} {\bibfnamefont {S.}~\bibnamefont {Nolte}}, \bibinfo {author} {\bibfnamefont {M.}~\bibnamefont {Segev}},\ and\ \bibinfo {author} {\bibfnamefont {A.}~\bibnamefont {Szameit}},\ }\bibfield  {title} {\bibinfo {title} {Photonic floquet topological insulators},\ }\href@noop {} {\bibfield  {journal} {\bibinfo  {journal} {Nature}\ }\textbf {\bibinfo {volume} {496}},\ \bibinfo {pages} {196} (\bibinfo {year} {2013})}\BibitemShut {NoStop}%
\bibitem [{\citenamefont {Hafezi}\ \emph {et~al.}(2013)\citenamefont {Hafezi}, \citenamefont {Mittal}, \citenamefont {Fan}, \citenamefont {Migdall},\ and\ \citenamefont {Taylor}}]{hafezi2013imaging}%
  \BibitemOpen
  \bibfield  {author} {\bibinfo {author} {\bibfnamefont {M.}~\bibnamefont {Hafezi}}, \bibinfo {author} {\bibfnamefont {S.}~\bibnamefont {Mittal}}, \bibinfo {author} {\bibfnamefont {J.}~\bibnamefont {Fan}}, \bibinfo {author} {\bibfnamefont {A.}~\bibnamefont {Migdall}},\ and\ \bibinfo {author} {\bibfnamefont {J.}~\bibnamefont {Taylor}},\ }\bibfield  {title} {\bibinfo {title} {Imaging topological edge states in silicon photonics},\ }\href@noop {} {\bibfield  {journal} {\bibinfo  {journal} {Nature Photonics}\ }\textbf {\bibinfo {volume} {7}},\ \bibinfo {pages} {1001} (\bibinfo {year} {2013})}\BibitemShut {NoStop}%
\bibitem [{\citenamefont {Jin}\ \emph {et~al.}(2025)\citenamefont {Jin}, \citenamefont {He}, \citenamefont {Lu}, \citenamefont {Chang}, \citenamefont {Shang}, \citenamefont {Bowers}, \citenamefont {Mele},\ and\ \citenamefont {Zhen}}]{jin2025towards}%
  \BibitemOpen
  \bibfield  {author} {\bibinfo {author} {\bibfnamefont {J.}~\bibnamefont {Jin}}, \bibinfo {author} {\bibfnamefont {L.}~\bibnamefont {He}}, \bibinfo {author} {\bibfnamefont {J.}~\bibnamefont {Lu}}, \bibinfo {author} {\bibfnamefont {L.}~\bibnamefont {Chang}}, \bibinfo {author} {\bibfnamefont {C.}~\bibnamefont {Shang}}, \bibinfo {author} {\bibfnamefont {J.~E.}\ \bibnamefont {Bowers}}, \bibinfo {author} {\bibfnamefont {E.~J.}\ \bibnamefont {Mele}},\ and\ \bibinfo {author} {\bibfnamefont {B.}~\bibnamefont {Zhen}},\ }\bibfield  {title} {\bibinfo {title} {Towards floquet chern insulators of light},\ }\href@noop {} {\bibfield  {journal} {\bibinfo  {journal} {Nature Nanotechnology}\ }\textbf {\bibinfo {volume} {20}},\ \bibinfo {pages} {1574} (\bibinfo {year} {2025})}\BibitemShut {NoStop}%
\bibitem [{\citenamefont {Berry}(1984)}]{berry1984quantal}%
  \BibitemOpen
  \bibfield  {author} {\bibinfo {author} {\bibfnamefont {M.~V.}\ \bibnamefont {Berry}},\ }\bibfield  {title} {\bibinfo {title} {Quantal phase factors accompanying adiabatic changes},\ }\href@noop {} {\bibfield  {journal} {\bibinfo  {journal} {Proc. R. Soc. Lond. A}\ }\textbf {\bibinfo {volume} {392}},\ \bibinfo {pages} {45} (\bibinfo {year} {1984})}\BibitemShut {NoStop}%
\bibitem [{\citenamefont {Armitage}\ \emph {et~al.}(2018)\citenamefont {Armitage}, \citenamefont {Mele},\ and\ \citenamefont {Vishwanath}}]{armitage2018weyl}%
  \BibitemOpen
  \bibfield  {author} {\bibinfo {author} {\bibfnamefont {N.~P.}\ \bibnamefont {Armitage}}, \bibinfo {author} {\bibfnamefont {E.~J.}\ \bibnamefont {Mele}},\ and\ \bibinfo {author} {\bibfnamefont {A.}~\bibnamefont {Vishwanath}},\ }\bibfield  {title} {\bibinfo {title} {Weyl and dirac semimetals in three-dimensional solids},\ }\href@noop {} {\bibfield  {journal} {\bibinfo  {journal} {Reviews of Modern Physics}\ }\textbf {\bibinfo {volume} {90}},\ \bibinfo {pages} {015001} (\bibinfo {year} {2018})}\BibitemShut {NoStop}%
\bibitem [{\citenamefont {Lu}\ \emph {et~al.}(2013)\citenamefont {Lu}, \citenamefont {Fu}, \citenamefont {Joannopoulos},\ and\ \citenamefont {Solja{\v{c}}i{\'c}}}]{lu2013weyl}%
  \BibitemOpen
  \bibfield  {author} {\bibinfo {author} {\bibfnamefont {L.}~\bibnamefont {Lu}}, \bibinfo {author} {\bibfnamefont {L.}~\bibnamefont {Fu}}, \bibinfo {author} {\bibfnamefont {J.~D.}\ \bibnamefont {Joannopoulos}},\ and\ \bibinfo {author} {\bibfnamefont {M.}~\bibnamefont {Solja{\v{c}}i{\'c}}},\ }\bibfield  {title} {\bibinfo {title} {Weyl points and line nodes in gyroid photonic crystals},\ }\href@noop {} {\bibfield  {journal} {\bibinfo  {journal} {Nature photonics}\ }\textbf {\bibinfo {volume} {7}},\ \bibinfo {pages} {294} (\bibinfo {year} {2013})}\BibitemShut {NoStop}%
\bibitem [{\citenamefont {Noh}\ \emph {et~al.}(2017)\citenamefont {Noh}, \citenamefont {Huang}, \citenamefont {Leykam}, \citenamefont {Chong}, \citenamefont {Chen},\ and\ \citenamefont {Rechtsman}}]{noh2017experimental}%
  \BibitemOpen
  \bibfield  {author} {\bibinfo {author} {\bibfnamefont {J.}~\bibnamefont {Noh}}, \bibinfo {author} {\bibfnamefont {S.}~\bibnamefont {Huang}}, \bibinfo {author} {\bibfnamefont {D.}~\bibnamefont {Leykam}}, \bibinfo {author} {\bibfnamefont {Y.~D.}\ \bibnamefont {Chong}}, \bibinfo {author} {\bibfnamefont {K.~P.}\ \bibnamefont {Chen}},\ and\ \bibinfo {author} {\bibfnamefont {M.~C.}\ \bibnamefont {Rechtsman}},\ }\bibfield  {title} {\bibinfo {title} {Experimental observation of optical {Weyl} points and fermi arc-like surface states},\ }\href@noop {} {\bibfield  {journal} {\bibinfo  {journal} {Nat. Phys.}\ }\textbf {\bibinfo {volume} {13}},\ \bibinfo {pages} {611} (\bibinfo {year} {2017})}\BibitemShut {NoStop}%
\bibitem [{\citenamefont {Vaidya}\ \emph {et~al.}(2020)\citenamefont {Vaidya}, \citenamefont {Noh}, \citenamefont {Cerjan}, \citenamefont {J{\"o}rg}, \citenamefont {Von~Freymann},\ and\ \citenamefont {Rechtsman}}]{vaidya2020observation}%
  \BibitemOpen
  \bibfield  {author} {\bibinfo {author} {\bibfnamefont {S.}~\bibnamefont {Vaidya}}, \bibinfo {author} {\bibfnamefont {J.}~\bibnamefont {Noh}}, \bibinfo {author} {\bibfnamefont {A.}~\bibnamefont {Cerjan}}, \bibinfo {author} {\bibfnamefont {C.}~\bibnamefont {J{\"o}rg}}, \bibinfo {author} {\bibfnamefont {G.}~\bibnamefont {Von~Freymann}},\ and\ \bibinfo {author} {\bibfnamefont {M.~C.}\ \bibnamefont {Rechtsman}},\ }\bibfield  {title} {\bibinfo {title} {Observation of a charge-2 photonic {Weyl} point in the infrared},\ }\href@noop {} {\bibfield  {journal} {\bibinfo  {journal} {Phys. Rev. Lett.}\ }\textbf {\bibinfo {volume} {125}},\ \bibinfo {pages} {253902} (\bibinfo {year} {2020})}\BibitemShut {NoStop}%
\bibitem [{\citenamefont {J{\"o}rg}\ \emph {et~al.}(2022)\citenamefont {J{\"o}rg}, \citenamefont {Vaidya}, \citenamefont {Noh}, \citenamefont {Cerjan}, \citenamefont {Augustine}, \citenamefont {von Freymann},\ and\ \citenamefont {Rechtsman}}]{jorg2022observation}%
  \BibitemOpen
  \bibfield  {author} {\bibinfo {author} {\bibfnamefont {C.}~\bibnamefont {J{\"o}rg}}, \bibinfo {author} {\bibfnamefont {S.}~\bibnamefont {Vaidya}}, \bibinfo {author} {\bibfnamefont {J.}~\bibnamefont {Noh}}, \bibinfo {author} {\bibfnamefont {A.}~\bibnamefont {Cerjan}}, \bibinfo {author} {\bibfnamefont {S.}~\bibnamefont {Augustine}}, \bibinfo {author} {\bibfnamefont {G.}~\bibnamefont {von Freymann}},\ and\ \bibinfo {author} {\bibfnamefont {M.~C.}\ \bibnamefont {Rechtsman}},\ }\bibfield  {title} {\bibinfo {title} {Observation of quadratic (charge-2) {Weyl} point splitting in near-infrared photonic crystals},\ }\href@noop {} {\bibfield  {journal} {\bibinfo  {journal} {Laser Photonics Rev.}\ }\textbf {\bibinfo {volume} {16}},\ \bibinfo {pages} {2100452} (\bibinfo {year} {2022})}\BibitemShut {NoStop}%
\bibitem [{\citenamefont {Pixley}\ \emph {et~al.}(2018)\citenamefont {Pixley}, \citenamefont {Wilson}, \citenamefont {Huse},\ and\ \citenamefont {Gopalakrishnan}}]{Sarang2018Weyl}%
  \BibitemOpen
  \bibfield  {author} {\bibinfo {author} {\bibfnamefont {J.~H.}\ \bibnamefont {Pixley}}, \bibinfo {author} {\bibfnamefont {J.~H.}\ \bibnamefont {Wilson}}, \bibinfo {author} {\bibfnamefont {D.~A.}\ \bibnamefont {Huse}},\ and\ \bibinfo {author} {\bibfnamefont {S.}~\bibnamefont {Gopalakrishnan}},\ }\bibfield  {title} {\bibinfo {title} {Weyl semimetal to metal phase transitions driven by quasiperiodic potentials},\ }\href {https://doi.org/10.1103/PhysRevLett.120.207604} {\bibfield  {journal} {\bibinfo  {journal} {Phys. Rev. Lett.}\ }\textbf {\bibinfo {volume} {120}},\ \bibinfo {pages} {207604} (\bibinfo {year} {2018})}\BibitemShut {NoStop}%
\bibitem [{\citenamefont {Castro~Neto}\ \emph {et~al.}(2009)\citenamefont {Castro~Neto}, \citenamefont {Guinea}, \citenamefont {Peres}, \citenamefont {Novoselov},\ and\ \citenamefont {Geim}}]{Castro2009TheGraphene}%
  \BibitemOpen
  \bibfield  {author} {\bibinfo {author} {\bibfnamefont {A.~H.}\ \bibnamefont {Castro~Neto}}, \bibinfo {author} {\bibfnamefont {F.}~\bibnamefont {Guinea}}, \bibinfo {author} {\bibfnamefont {N.~M.~R.}\ \bibnamefont {Peres}}, \bibinfo {author} {\bibfnamefont {K.~S.}\ \bibnamefont {Novoselov}},\ and\ \bibinfo {author} {\bibfnamefont {A.~K.}\ \bibnamefont {Geim}},\ }\bibfield  {title} {\bibinfo {title} {The electronic properties of graphene},\ }\href {https://doi.org/10.1103/RevModPhys.81.109} {\bibfield  {journal} {\bibinfo  {journal} {Rev. Mod. Phys.}\ }\textbf {\bibinfo {volume} {81}},\ \bibinfo {pages} {109} (\bibinfo {year} {2009})}\BibitemShut {NoStop}%
\bibitem [{\citenamefont {Xu}\ \emph {et~al.}(2015)\citenamefont {Xu}, \citenamefont {Belopolski}, \citenamefont {Alidoust}, \citenamefont {Neupane}, \citenamefont {Bian}, \citenamefont {Zhang}, \citenamefont {Sankar}, \citenamefont {Chang}, \citenamefont {Yuan}, \citenamefont {Lee}, \citenamefont {Huang}, \citenamefont {Zheng}, \citenamefont {Ma}, \citenamefont {Sanchez}, \citenamefont {Wang}, \citenamefont {Bansil}, \citenamefont {Chou}, \citenamefont {Shibayev}, \citenamefont {Lin}, \citenamefont {Jia},\ and\ \citenamefont {Hasan}}]{Xu2015DiscoveryScience}%
  \BibitemOpen
  \bibfield  {author} {\bibinfo {author} {\bibfnamefont {S.-Y.}\ \bibnamefont {Xu}}, \bibinfo {author} {\bibfnamefont {I.}~\bibnamefont {Belopolski}}, \bibinfo {author} {\bibfnamefont {N.}~\bibnamefont {Alidoust}}, \bibinfo {author} {\bibfnamefont {M.}~\bibnamefont {Neupane}}, \bibinfo {author} {\bibfnamefont {G.}~\bibnamefont {Bian}}, \bibinfo {author} {\bibfnamefont {C.}~\bibnamefont {Zhang}}, \bibinfo {author} {\bibfnamefont {R.}~\bibnamefont {Sankar}}, \bibinfo {author} {\bibfnamefont {G.}~\bibnamefont {Chang}}, \bibinfo {author} {\bibfnamefont {Z.}~\bibnamefont {Yuan}}, \bibinfo {author} {\bibfnamefont {C.-C.}\ \bibnamefont {Lee}}, \bibinfo {author} {\bibfnamefont {S.-M.}\ \bibnamefont {Huang}}, \bibinfo {author} {\bibfnamefont {H.}~\bibnamefont {Zheng}}, \bibinfo {author} {\bibfnamefont {J.}~\bibnamefont {Ma}}, \bibinfo {author} {\bibfnamefont {D.~S.}\ \bibnamefont {Sanchez}}, \bibinfo {author} {\bibfnamefont {B.}~\bibnamefont {Wang}}, \bibinfo {author} {\bibfnamefont {A.}~\bibnamefont {Bansil}},
  \bibinfo {author} {\bibfnamefont {F.}~\bibnamefont {Chou}}, \bibinfo {author} {\bibfnamefont {P.~P.}\ \bibnamefont {Shibayev}}, \bibinfo {author} {\bibfnamefont {H.}~\bibnamefont {Lin}}, \bibinfo {author} {\bibfnamefont {S.}~\bibnamefont {Jia}},\ and\ \bibinfo {author} {\bibfnamefont {M.~Z.}\ \bibnamefont {Hasan}},\ }\bibfield  {title} {\bibinfo {title} {Discovery of a $\rm{W}eyl$ fermion semimetal and topological $\rm{F}ermi$ arcs},\ }\href {https://doi.org/10.1126/science.aaa9297} {\bibfield  {journal} {\bibinfo  {journal} {Science}\ }\textbf {\bibinfo {volume} {349}},\ \bibinfo {pages} {613} (\bibinfo {year} {2015})}\BibitemShut {NoStop}%
\bibitem [{\citenamefont {Lv}\ \emph {et~al.}(2015)\citenamefont {Lv}, \citenamefont {Xu}, \citenamefont {Weng}, \citenamefont {Ma}, \citenamefont {Richard}, \citenamefont {Huang}, \citenamefont {Zhao}, \citenamefont {Chen}, \citenamefont {Matt}, \citenamefont {Bisti}, \citenamefont {Strocov}, \citenamefont {Mesot}, \citenamefont {Fang}, \citenamefont {Dai}, \citenamefont {Qian}, \citenamefont {Shi},\ and\ \citenamefont {Ding}}]{Lv2015Observation}%
  \BibitemOpen
  \bibfield  {author} {\bibinfo {author} {\bibfnamefont {B.~Q.}\ \bibnamefont {Lv}}, \bibinfo {author} {\bibfnamefont {N.}~\bibnamefont {Xu}}, \bibinfo {author} {\bibfnamefont {H.~M.}\ \bibnamefont {Weng}}, \bibinfo {author} {\bibfnamefont {J.~Z.}\ \bibnamefont {Ma}}, \bibinfo {author} {\bibfnamefont {P.}~\bibnamefont {Richard}}, \bibinfo {author} {\bibfnamefont {X.~C.}\ \bibnamefont {Huang}}, \bibinfo {author} {\bibfnamefont {L.~X.}\ \bibnamefont {Zhao}}, \bibinfo {author} {\bibfnamefont {G.~F.}\ \bibnamefont {Chen}}, \bibinfo {author} {\bibfnamefont {C.~E.}\ \bibnamefont {Matt}}, \bibinfo {author} {\bibfnamefont {F.}~\bibnamefont {Bisti}}, \bibinfo {author} {\bibfnamefont {V.~N.}\ \bibnamefont {Strocov}}, \bibinfo {author} {\bibfnamefont {J.}~\bibnamefont {Mesot}}, \bibinfo {author} {\bibfnamefont {Z.}~\bibnamefont {Fang}}, \bibinfo {author} {\bibfnamefont {X.}~\bibnamefont {Dai}}, \bibinfo {author} {\bibfnamefont {T.}~\bibnamefont {Qian}}, \bibinfo {author} {\bibfnamefont {M.}~\bibnamefont {Shi}},\ and\
  \bibinfo {author} {\bibfnamefont {H.}~\bibnamefont {Ding}},\ }\bibfield  {title} {\bibinfo {title} {Observation of $\rm{W}eyl$ nodes in $\rm{TaAs}$},\ }\href {https://doi.org/10.1038/Nphys3426} {\bibfield  {journal} {\bibinfo  {journal} {Nat. Phys.}\ }\textbf {\bibinfo {volume} {11}},\ \bibinfo {pages} {724} (\bibinfo {year} {2015})}\BibitemShut {NoStop}%
\bibitem [{\citenamefont {Fruchart}\ \emph {et~al.}(2018)\citenamefont {Fruchart}, \citenamefont {Jeon}, \citenamefont {Hur}, \citenamefont {Cheianov}, \citenamefont {Wiesner},\ and\ \citenamefont {Vitelli}}]{fruchart2018soft}%
  \BibitemOpen
  \bibfield  {author} {\bibinfo {author} {\bibfnamefont {M.}~\bibnamefont {Fruchart}}, \bibinfo {author} {\bibfnamefont {S.-Y.}\ \bibnamefont {Jeon}}, \bibinfo {author} {\bibfnamefont {K.}~\bibnamefont {Hur}}, \bibinfo {author} {\bibfnamefont {V.}~\bibnamefont {Cheianov}}, \bibinfo {author} {\bibfnamefont {U.}~\bibnamefont {Wiesner}},\ and\ \bibinfo {author} {\bibfnamefont {V.}~\bibnamefont {Vitelli}},\ }\bibfield  {title} {\bibinfo {title} {Soft self-assembly of weyl materials for light and sound},\ }\href@noop {} {\bibfield  {journal} {\bibinfo  {journal} {Proceedings of the National Academy of Sciences}\ }\textbf {\bibinfo {volume} {115}},\ \bibinfo {pages} {E3655} (\bibinfo {year} {2018})}\BibitemShut {NoStop}%
\bibitem [{\citenamefont {Hajduk}\ \emph {et~al.}(1994)\citenamefont {Hajduk}, \citenamefont {Harper}, \citenamefont {Gruner}, \citenamefont {Honeker}, \citenamefont {Kim}, \citenamefont {Thomas},\ and\ \citenamefont {Fetters}}]{hajduk1994gyroid}%
  \BibitemOpen
  \bibfield  {author} {\bibinfo {author} {\bibfnamefont {D.~A.}\ \bibnamefont {Hajduk}}, \bibinfo {author} {\bibfnamefont {P.~E.}\ \bibnamefont {Harper}}, \bibinfo {author} {\bibfnamefont {S.~M.}\ \bibnamefont {Gruner}}, \bibinfo {author} {\bibfnamefont {C.~C.}\ \bibnamefont {Honeker}}, \bibinfo {author} {\bibfnamefont {G.}~\bibnamefont {Kim}}, \bibinfo {author} {\bibfnamefont {E.~L.}\ \bibnamefont {Thomas}},\ and\ \bibinfo {author} {\bibfnamefont {L.~J.}\ \bibnamefont {Fetters}},\ }\bibfield  {title} {\bibinfo {title} {The gyroid: a new equilibrium morphology in weakly segregated diblock copolymers},\ }\href@noop {} {\bibfield  {journal} {\bibinfo  {journal} {Macromolecules}\ }\textbf {\bibinfo {volume} {27}},\ \bibinfo {pages} {4063} (\bibinfo {year} {1994})}\BibitemShut {NoStop}%
\bibitem [{\citenamefont {Pusey}\ and\ \citenamefont {Van~Megen}(1986)}]{pusey_phase_1986}%
  \BibitemOpen
  \bibfield  {author} {\bibinfo {author} {\bibfnamefont {P.~N.}\ \bibnamefont {Pusey}}\ and\ \bibinfo {author} {\bibfnamefont {W.}~\bibnamefont {Van~Megen}},\ }\bibfield  {title} {\bibinfo {title} {Phase behaviour of concentrated suspensions of nearly hard colloidal spheres},\ }\href {https://doi.org/10.1038/320340a0} {\bibfield  {journal} {\bibinfo  {journal} {Nature}\ }\textbf {\bibinfo {volume} {320}},\ \bibinfo {pages} {340} (\bibinfo {year} {1986})}\BibitemShut {NoStop}%
\bibitem [{\citenamefont {Yablonovitch}(1987)}]{yablonovitch_inhibited_1987}%
  \BibitemOpen
  \bibfield  {author} {\bibinfo {author} {\bibfnamefont {E.}~\bibnamefont {Yablonovitch}},\ }\bibfield  {title} {\bibinfo {title} {Inhibited spontaneous emission in solid-state physics and electronics},\ }\href {https://doi.org/10.1103/PhysRevLett.58.2059} {\bibfield  {journal} {\bibinfo  {journal} {Phys. Rev. Lett.}\ }\textbf {\bibinfo {volume} {58}},\ \bibinfo {pages} {2059} (\bibinfo {year} {1987})}\BibitemShut {NoStop}%
\bibitem [{\citenamefont {John}(1987)}]{John1987Strong}%
  \BibitemOpen
  \bibfield  {author} {\bibinfo {author} {\bibfnamefont {S.}~\bibnamefont {John}},\ }\bibfield  {title} {\bibinfo {title} {Strong localization of photons in certain disordered dielectric superlattices},\ }\href {https://doi.org/10.1103/PhysRevLett.58.2486} {\bibfield  {journal} {\bibinfo  {journal} {Phys. Rev. Lett.}\ }\textbf {\bibinfo {volume} {58}},\ \bibinfo {pages} {2486} (\bibinfo {year} {1987})}\BibitemShut {NoStop}%
\bibitem [{\citenamefont {Yablonovitch}(1993)}]{Yablonovitch1993Photonic}%
  \BibitemOpen
  \bibfield  {author} {\bibinfo {author} {\bibfnamefont {E.}~\bibnamefont {Yablonovitch}},\ }\bibfield  {title} {\bibinfo {title} {Photonic band-gap structures},\ }\href {https://doi.org/10.1364/JOSAB.10.000283} {\bibfield  {journal} {\bibinfo  {journal} {J. Opt. Soc. Am. B}\ }\textbf {\bibinfo {volume} {10}},\ \bibinfo {pages} {283} (\bibinfo {year} {1993})}\BibitemShut {NoStop}%
\bibitem [{\citenamefont {Vlasov}\ \emph {et~al.}(2001)\citenamefont {Vlasov}, \citenamefont {Bo}, \citenamefont {Sturm},\ and\ \citenamefont {Norris}}]{vlasov2001chip}%
  \BibitemOpen
  \bibfield  {author} {\bibinfo {author} {\bibfnamefont {Y.~A.}\ \bibnamefont {Vlasov}}, \bibinfo {author} {\bibfnamefont {X.-Z.}\ \bibnamefont {Bo}}, \bibinfo {author} {\bibfnamefont {J.~C.}\ \bibnamefont {Sturm}},\ and\ \bibinfo {author} {\bibfnamefont {D.~J.}\ \bibnamefont {Norris}},\ }\bibfield  {title} {\bibinfo {title} {On-chip natural assembly of silicon photonic bandgap crystals},\ }\href@noop {} {\bibfield  {journal} {\bibinfo  {journal} {Nature}\ }\textbf {\bibinfo {volume} {414}},\ \bibinfo {pages} {289} (\bibinfo {year} {2001})}\BibitemShut {NoStop}%
\bibitem [{\citenamefont {Ho}\ \emph {et~al.}(1990)\citenamefont {Ho}, \citenamefont {Chan},\ and\ \citenamefont {Soukoulis}}]{Ho1990}%
  \BibitemOpen
  \bibfield  {author} {\bibinfo {author} {\bibfnamefont {K.~M.}\ \bibnamefont {Ho}}, \bibinfo {author} {\bibfnamefont {C.~T.}\ \bibnamefont {Chan}},\ and\ \bibinfo {author} {\bibfnamefont {C.~M.}\ \bibnamefont {Soukoulis}},\ }\bibfield  {title} {\bibinfo {title} {Existence of a photonic gap in periodic dielectric structures},\ }\href {https://doi.org/10.1103/PhysRevLett.65.3152} {\bibfield  {journal} {\bibinfo  {journal} {Phys. Rev. Lett.}\ }\textbf {\bibinfo {volume} {65}},\ \bibinfo {pages} {3152} (\bibinfo {year} {1990})}\BibitemShut {NoStop}%
\bibitem [{\citenamefont {He}\ \emph {et~al.}(2020)\citenamefont {He}, \citenamefont {Gales}, \citenamefont {Ducrot}, \citenamefont {Gong}, \citenamefont {Yi}, \citenamefont {Sacanna},\ and\ \citenamefont {Pine}}]{he2020colloidal}%
  \BibitemOpen
  \bibfield  {author} {\bibinfo {author} {\bibfnamefont {M.}~\bibnamefont {He}}, \bibinfo {author} {\bibfnamefont {J.~P.}\ \bibnamefont {Gales}}, \bibinfo {author} {\bibfnamefont {{\'E}.}~\bibnamefont {Ducrot}}, \bibinfo {author} {\bibfnamefont {Z.}~\bibnamefont {Gong}}, \bibinfo {author} {\bibfnamefont {G.-R.}\ \bibnamefont {Yi}}, \bibinfo {author} {\bibfnamefont {S.}~\bibnamefont {Sacanna}},\ and\ \bibinfo {author} {\bibfnamefont {D.~J.}\ \bibnamefont {Pine}},\ }\bibfield  {title} {\bibinfo {title} {Colloidal diamond},\ }\href@noop {} {\bibfield  {journal} {\bibinfo  {journal} {Nature}\ }\textbf {\bibinfo {volume} {585}},\ \bibinfo {pages} {524} (\bibinfo {year} {2020})}\BibitemShut {NoStop}%
\bibitem [{\citenamefont {Kalsin}\ \emph {et~al.}(2006)\citenamefont {Kalsin}, \citenamefont {Fialkowski}, \citenamefont {Paszewski}, \citenamefont {Smoukov}, \citenamefont {Bishop},\ and\ \citenamefont {Grzybowski}}]{kalsin2006electrostatic}%
  \BibitemOpen
  \bibfield  {author} {\bibinfo {author} {\bibfnamefont {A.~M.}\ \bibnamefont {Kalsin}}, \bibinfo {author} {\bibfnamefont {M.}~\bibnamefont {Fialkowski}}, \bibinfo {author} {\bibfnamefont {M.}~\bibnamefont {Paszewski}}, \bibinfo {author} {\bibfnamefont {S.~K.}\ \bibnamefont {Smoukov}}, \bibinfo {author} {\bibfnamefont {K.~J.}\ \bibnamefont {Bishop}},\ and\ \bibinfo {author} {\bibfnamefont {B.~A.}\ \bibnamefont {Grzybowski}},\ }\bibfield  {title} {\bibinfo {title} {Electrostatic self-assembly of binary nanoparticle crystals with a diamond-like lattice},\ }\href@noop {} {\bibfield  {journal} {\bibinfo  {journal} {science}\ }\textbf {\bibinfo {volume} {312}},\ \bibinfo {pages} {420} (\bibinfo {year} {2006})}\BibitemShut {NoStop}%
\bibitem [{\citenamefont {Liu}\ \emph {et~al.}(2016)\citenamefont {Liu}, \citenamefont {Tagawa}, \citenamefont {Xin}, \citenamefont {Wang}, \citenamefont {Emamy}, \citenamefont {Li}, \citenamefont {Yager}, \citenamefont {Starr}, \citenamefont {Tkachenko},\ and\ \citenamefont {Gang}}]{liu2016diamond}%
  \BibitemOpen
  \bibfield  {author} {\bibinfo {author} {\bibfnamefont {W.}~\bibnamefont {Liu}}, \bibinfo {author} {\bibfnamefont {M.}~\bibnamefont {Tagawa}}, \bibinfo {author} {\bibfnamefont {H.~L.}\ \bibnamefont {Xin}}, \bibinfo {author} {\bibfnamefont {T.}~\bibnamefont {Wang}}, \bibinfo {author} {\bibfnamefont {H.}~\bibnamefont {Emamy}}, \bibinfo {author} {\bibfnamefont {H.}~\bibnamefont {Li}}, \bibinfo {author} {\bibfnamefont {K.~G.}\ \bibnamefont {Yager}}, \bibinfo {author} {\bibfnamefont {F.~W.}\ \bibnamefont {Starr}}, \bibinfo {author} {\bibfnamefont {A.~V.}\ \bibnamefont {Tkachenko}},\ and\ \bibinfo {author} {\bibfnamefont {O.}~\bibnamefont {Gang}},\ }\bibfield  {title} {\bibinfo {title} {Diamond family of nanoparticle superlattices},\ }\href@noop {} {\bibfield  {journal} {\bibinfo  {journal} {Science}\ }\textbf {\bibinfo {volume} {351}},\ \bibinfo {pages} {582} (\bibinfo {year} {2016})}\BibitemShut {NoStop}%
\bibitem [{\citenamefont {Bragg}\ and\ \citenamefont {Bragg}(1913)}]{bragg1913structure}%
  \BibitemOpen
  \bibfield  {author} {\bibinfo {author} {\bibfnamefont {W.~H.}\ \bibnamefont {Bragg}}\ and\ \bibinfo {author} {\bibfnamefont {W.~L.}\ \bibnamefont {Bragg}},\ }\bibfield  {title} {\bibinfo {title} {The structure of the diamond},\ }\href@noop {} {\bibfield  {journal} {\bibinfo  {journal} {Nature}\ }\textbf {\bibinfo {volume} {91}},\ \bibinfo {pages} {557} (\bibinfo {year} {1913})}\BibitemShut {NoStop}%
\bibitem [{\citenamefont {Wang}\ \emph {et~al.}(2012)\citenamefont {Wang}, \citenamefont {Wang}, \citenamefont {Breed}, \citenamefont {Manoharan}, \citenamefont {Feng}, \citenamefont {Hollingsworth}, \citenamefont {Weck},\ and\ \citenamefont {Pine}}]{wang2012colloids}%
  \BibitemOpen
  \bibfield  {author} {\bibinfo {author} {\bibfnamefont {Y.}~\bibnamefont {Wang}}, \bibinfo {author} {\bibfnamefont {Y.}~\bibnamefont {Wang}}, \bibinfo {author} {\bibfnamefont {D.~R.}\ \bibnamefont {Breed}}, \bibinfo {author} {\bibfnamefont {V.~N.}\ \bibnamefont {Manoharan}}, \bibinfo {author} {\bibfnamefont {L.}~\bibnamefont {Feng}}, \bibinfo {author} {\bibfnamefont {A.~D.}\ \bibnamefont {Hollingsworth}}, \bibinfo {author} {\bibfnamefont {M.}~\bibnamefont {Weck}},\ and\ \bibinfo {author} {\bibfnamefont {D.~J.}\ \bibnamefont {Pine}},\ }\bibfield  {title} {\bibinfo {title} {Colloids with valence and specific directional bonding},\ }\href@noop {} {\bibfield  {journal} {\bibinfo  {journal} {Nature}\ }\textbf {\bibinfo {volume} {491}},\ \bibinfo {pages} {51} (\bibinfo {year} {2012})}\BibitemShut {NoStop}%
\bibitem [{\citenamefont {Gales}\ \emph {et~al.}(2025)\citenamefont {Gales}, \citenamefont {Kim}, \citenamefont {Hocky},\ and\ \citenamefont {Pine}}]{gales2025crystallization}%
  \BibitemOpen
  \bibfield  {author} {\bibinfo {author} {\bibfnamefont {J.~P.}\ \bibnamefont {Gales}}, \bibinfo {author} {\bibfnamefont {M.~J.}\ \bibnamefont {Kim}}, \bibinfo {author} {\bibfnamefont {G.~M.}\ \bibnamefont {Hocky}},\ and\ \bibinfo {author} {\bibfnamefont {D.~J.}\ \bibnamefont {Pine}},\ }\bibfield  {title} {\bibinfo {title} {Crystallization of non-convex colloids: the roles of particle shape and entropy},\ }\href@noop {} {\bibfield  {journal} {\bibinfo  {journal} {Soft Matter}\ }\textbf {\bibinfo {volume} {21}},\ \bibinfo {pages} {7021} (\bibinfo {year} {2025})}\BibitemShut {NoStop}%
\bibitem [{\citenamefont {Johnson}\ and\ \citenamefont {Joannopoulos}(2001)}]{johnson2001block}%
  \BibitemOpen
  \bibfield  {author} {\bibinfo {author} {\bibfnamefont {S.~G.}\ \bibnamefont {Johnson}}\ and\ \bibinfo {author} {\bibfnamefont {J.~D.}\ \bibnamefont {Joannopoulos}},\ }\bibfield  {title} {\bibinfo {title} {Block-iterative frequency-domain methods for maxwell’s equations in a planewave basis},\ }\href@noop {} {\bibfield  {journal} {\bibinfo  {journal} {Optics express}\ }\textbf {\bibinfo {volume} {8}},\ \bibinfo {pages} {173} (\bibinfo {year} {2001})}\BibitemShut {NoStop}%
\bibitem [{\citenamefont {Yu}\ \emph {et~al.}(2022)\citenamefont {Yu}, \citenamefont {Zhang}, \citenamefont {Liu}, \citenamefont {Wu}, \citenamefont {Li}, \citenamefont {Zhang}, \citenamefont {Yang},\ and\ \citenamefont {Yao}}]{yu2022encyclopedia}%
  \BibitemOpen
  \bibfield  {author} {\bibinfo {author} {\bibfnamefont {Z.-M.}\ \bibnamefont {Yu}}, \bibinfo {author} {\bibfnamefont {Z.}~\bibnamefont {Zhang}}, \bibinfo {author} {\bibfnamefont {G.-B.}\ \bibnamefont {Liu}}, \bibinfo {author} {\bibfnamefont {W.}~\bibnamefont {Wu}}, \bibinfo {author} {\bibfnamefont {X.-P.}\ \bibnamefont {Li}}, \bibinfo {author} {\bibfnamefont {R.-W.}\ \bibnamefont {Zhang}}, \bibinfo {author} {\bibfnamefont {S.~A.}\ \bibnamefont {Yang}},\ and\ \bibinfo {author} {\bibfnamefont {Y.}~\bibnamefont {Yao}},\ }\bibfield  {title} {\bibinfo {title} {Encyclopedia of emergent particles in three-dimensional crystals},\ }\href@noop {} {\bibfield  {journal} {\bibinfo  {journal} {Sci. Bull.}\ }\textbf {\bibinfo {volume} {67}},\ \bibinfo {pages} {375} (\bibinfo {year} {2022})}\BibitemShut {NoStop}%
\bibitem [{\citenamefont {Resta}(2000)}]{resta2000manifestations}%
  \BibitemOpen
  \bibfield  {author} {\bibinfo {author} {\bibfnamefont {R.}~\bibnamefont {Resta}},\ }\bibfield  {title} {\bibinfo {title} {Manifestations of berry's phase in molecules and condensed matter},\ }\href@noop {} {\bibfield  {journal} {\bibinfo  {journal} {J. Phys.: Condens. Matter}\ }\textbf {\bibinfo {volume} {12}},\ \bibinfo {pages} {R107} (\bibinfo {year} {2000})}\BibitemShut {NoStop}%
\bibitem [{\citenamefont {Tidy3D}()}]{tidy3dfdtd}%
  \BibitemOpen
  \bibfield  {author} {\bibinfo {author} {\bibfnamefont {F.-C.}\ \bibnamefont {Tidy3D}},\ }\href@noop {} {\bibinfo {title} {Fdtd for electromagnetic simulation}}\BibitemShut {NoStop}%
\bibitem [{\citenamefont {Glaser}\ \emph {et~al.}(2015)\citenamefont {Glaser}, \citenamefont {Nguyen}, \citenamefont {Anderson}, \citenamefont {Lui}, \citenamefont {Spiga}, \citenamefont {Millan}, \citenamefont {Morse},\ and\ \citenamefont {Glotzer}}]{hoomd-blue}%
  \BibitemOpen
  \bibfield  {author} {\bibinfo {author} {\bibfnamefont {J.}~\bibnamefont {Glaser}}, \bibinfo {author} {\bibfnamefont {T.~D.}\ \bibnamefont {Nguyen}}, \bibinfo {author} {\bibfnamefont {J.~A.}\ \bibnamefont {Anderson}}, \bibinfo {author} {\bibfnamefont {P.}~\bibnamefont {Lui}}, \bibinfo {author} {\bibfnamefont {F.}~\bibnamefont {Spiga}}, \bibinfo {author} {\bibfnamefont {J.~A.}\ \bibnamefont {Millan}}, \bibinfo {author} {\bibfnamefont {D.~C.}\ \bibnamefont {Morse}},\ and\ \bibinfo {author} {\bibfnamefont {S.~C.}\ \bibnamefont {Glotzer}},\ }\bibfield  {title} {\bibinfo {title} {Strong scaling of general-purpose molecular dynamics simulations on {GPUs}},\ }\href {https://doi.org/10.1016/j.cpc.2015.02.028} {\bibfield  {journal} {\bibinfo  {journal} {Comput. Phys. Commun.}\ }\textbf {\bibinfo {volume} {192}},\ \bibinfo {pages} {97} (\bibinfo {year} {2015})}\BibitemShut {NoStop}%
\bibitem [{\citenamefont {Wang}\ \emph {et~al.}(2015)\citenamefont {Wang}, \citenamefont {Wang}, \citenamefont {Zheng}, \citenamefont {Ducrot}, \citenamefont {Lee}, \citenamefont {Yi}, \citenamefont {Weck},\ and\ \citenamefont {Pine}}]{wang_synthetic_2015}%
  \BibitemOpen
  \bibfield  {author} {\bibinfo {author} {\bibfnamefont {Y.}~\bibnamefont {Wang}}, \bibinfo {author} {\bibfnamefont {Y.}~\bibnamefont {Wang}}, \bibinfo {author} {\bibfnamefont {X.}~\bibnamefont {Zheng}}, \bibinfo {author} {\bibfnamefont {E.}~\bibnamefont {Ducrot}}, \bibinfo {author} {\bibfnamefont {M.-G.}\ \bibnamefont {Lee}}, \bibinfo {author} {\bibfnamefont {G.-R.}\ \bibnamefont {Yi}}, \bibinfo {author} {\bibfnamefont {M.}~\bibnamefont {Weck}},\ and\ \bibinfo {author} {\bibfnamefont {D.~J.}\ \bibnamefont {Pine}},\ }\bibfield  {title} {\bibinfo {title} {Synthetic strategies toward {DNA}-coated colloids that crystallize},\ }\href {https://doi.org/10.1021/jacs.5b06607} {\bibfield  {journal} {\bibinfo  {journal} {J. Am. Chem. Soc.}\ }\textbf {\bibinfo {volume} {137}},\ \bibinfo {pages} {10760} (\bibinfo {year} {2015})}\BibitemShut {NoStop}%
\bibitem [{\citenamefont {Wang}\ \emph {et~al.}(2020)\citenamefont {Wang}, \citenamefont {Ram{\'\i}rez-Hinestrosa}, \citenamefont {Dobnikar},\ and\ \citenamefont {Frenkel}}]{wang2020lennard}%
  \BibitemOpen
  \bibfield  {author} {\bibinfo {author} {\bibfnamefont {X.}~\bibnamefont {Wang}}, \bibinfo {author} {\bibfnamefont {S.}~\bibnamefont {Ram{\'\i}rez-Hinestrosa}}, \bibinfo {author} {\bibfnamefont {J.}~\bibnamefont {Dobnikar}},\ and\ \bibinfo {author} {\bibfnamefont {D.}~\bibnamefont {Frenkel}},\ }\bibfield  {title} {\bibinfo {title} {The lennard-jones potential: when (not) to use it},\ }\href@noop {} {\bibfield  {journal} {\bibinfo  {journal} {Phys. Chem. Chem. Phys.}\ }\textbf {\bibinfo {volume} {22}},\ \bibinfo {pages} {10624} (\bibinfo {year} {2020})}\BibitemShut {NoStop}%
\bibitem [{\citenamefont {Cui}\ \emph {et~al.}(2022)\citenamefont {Cui}, \citenamefont {Marbach}, \citenamefont {Zheng}, \citenamefont {Holmes-Cerfon},\ and\ \citenamefont {Pine}}]{cui2022comprehensive}%
  \BibitemOpen
  \bibfield  {author} {\bibinfo {author} {\bibfnamefont {F.}~\bibnamefont {Cui}}, \bibinfo {author} {\bibfnamefont {S.}~\bibnamefont {Marbach}}, \bibinfo {author} {\bibfnamefont {J.~A.}\ \bibnamefont {Zheng}}, \bibinfo {author} {\bibfnamefont {M.}~\bibnamefont {Holmes-Cerfon}},\ and\ \bibinfo {author} {\bibfnamefont {D.~J.}\ \bibnamefont {Pine}},\ }\bibfield  {title} {\bibinfo {title} {Comprehensive view of microscopic interactions between {DNA}-coated colloids},\ }\href@noop {} {\bibfield  {journal} {\bibinfo  {journal} {Nat. Commun.}\ }\textbf {\bibinfo {volume} {13}},\ \bibinfo {pages} {2304} (\bibinfo {year} {2022})}\BibitemShut {NoStop}%
\end{thebibliography}
\end{document}